\documentclass{rsauthor}
\usepackage{graphicx}

\begin{document}
\renewcommand\thesubsection{\thesection.\arabic{subsection}}
\title{ Enhanced signal-to-noise ratios in frog hearing can be achieved through
amplitude death  }
\author{Kang-Hun Ahn
\address{Department of Physics, Chungnam National University, Daejeon
305-764, Republic of Korea}}
\date{\today}
\abstract{ In the ear, hair cells transform mechanical stimuli into
neuronal signals
 with great sensitivity relying on certain active processes.
 Individual hair cell
bundles of non-mammals such as frogs and turtles are known to show
spontaneous oscillation. However hair bundles in vivo must be quiet
in the absence of stimuli, otherwise, the signal is drowned in
intrinsic noise. Thus, a certain mechanism is needed to exist in
order to suppress intrinsic noise. Here, through a model study of
elastically coupled hair bundles of bullfrog sacculi, we show that a
low stimulus threshold and a high signal-to-noise ratio (SNR) can be
achieved through the amplitude death phenomenon (the cessation of
spontaneous oscillations by coupling). This phenomenon occurs only
when the coupled hair bundles have inhomogeneous distribution, which
is likely to be the case in biological systems.
 We show that the SNR has
non-monotonic dependence on the mass of the overlying membrane, and
find out that the SNR has maximum value in the region of the
amplitude death. The low threshold of stimulus through amplitude
death may account for the experimentally observed high sensitivity
of frog sacculi in detecting vibration. The hair bundles' amplitude
death mechanism provides a smart engineering design for low-noise
amplification.
 }
 \maketitle

\section{Introduction}

The ear can actively amplify weak signals to achieve great
sensitivity and a wide dynamic range of hearing. Hair cells of the
vertebrate inner ear are the mechanotransducers which have been
proposed to amplify weak signals by generating active
forces\cite{crawford1985,howard1987}. While amplification in the
mammalian cochlea is widely believed to originate from the membrane
dynamics involving outer hair cell motility\cite{ashmore},
non-mammalian vertebrates lack outer hair cells. Nevertheless the
ear of lower vertebrates achieves acute
hearing\cite{narins,lewis1988,lewis01}. The exact mechanism is not
yet clearly known. Hair bundle motility probably underlies the
amplification process. Unlike mammalian hair cells, spontaneous
oscillations have been observed in individual hair cells of
turtles\cite{crawford1985} and frogs\cite{howard1987,martin2003}.
The spontaneous oscillations are believed to result from adaptation
dynamics driven by molecular motors in hair
bundles\cite{crawford1985,howard1987,martin2000}. The oscillation
may provide an amplification mechanism through a synchronization
process\cite{strogatz,pikovsky}, where the oscillation frequency is
locked to the frequency of the external stimulus.
 However, in recent
experiments under more natural conditions than previously studied,
frog hair bundles with an overlying membrane
  are found not to spontaneously oscillate, but are in fact quiescent\cite{strimbu2010,strimbu2012}.
Furthermore, one can imagine that the auditory neuron would receive
strong noisy signals from spontaneous
oscillations\cite{howard1987,martin2003,martin2000}, as their
magnitudes are about 20-50 nm which can cause significant changes in
the open probability of the mechanotransducer
channel\cite{martin2003,martin2000}.

If the spontaneous activity of the bundles is suppressed by their
overlying membrane, how does this activity contribute to frogs'
auditory transduction? Here we try to answer this question by
providing a low-noise amplification mechanism for the bullfrog
sacculus. Using numerical simulations of models of bullfrog sacculi,
we investigate the mechanotransduction properties of inhomogeneous
hair bundles with elastic coupling and mechanical loading. We show
that a low-noise amplification arises as a result of inter-bundle
interactions and hair bundle motility. A phenomenon that provides
for this low-noise amplification mechanism is amplitude death - the
cessation of oscillation due to the coupling of
oscillators\cite{aronson}. This intriguing phenomenon was first
noted in the 19th century by Rayleigh, who found that adjacent organ
pipes can suppress each other's sound\cite{rayleigh}.

Since the amplitude death is a universal phenomenon appearing when
any two or more different oscillators are coupled\cite{ryu,
pikovsky}, it has gained a great deal of attention in the physics
community. The required conditions for the amplitude death of the
oscillators are the coupling between them and their inhomogeneity.
The frog sacculus may satisfy these requirements. First, the hair
bundles of a frog's sacculus in vivo are coupled through the
otolithic membrane\cite{smotherman}. As for the inhomogeneity, the
hair bundles may have different dynamical properties. Experiments
report that some of the hair bundles show spontaneous oscillation
while the others remain quiescent\cite{howard1987,strimbu2010}. The
frequencies of spontaneous oscillations in the sacculus are randomly
distributed in a sacculus with a range of 5 Hz - 50
Hz\cite{ramunno-johnson}.

Noise is the natural constraint that limits the sensitivity of
sensory systems.
 To investigate the noise effect carefully, we develop a numerical
calculation method for thermal noise force.  In the absence of any
active process, according to the equipartition theorem, the average kinetic energy of a passive mechanical
sensor in thermal equilibrium is given by the thermal energy. This theorem is satisfied by the thermal
noise force. Equipped with the noise force, we simulate the dynamics
of hair bundles with an overlying membrane, and find that the
amplitude death phenomenon suppresses noise and enhances the signal
transmission. We find  that there exists an optimal value of the
mass of the overlying membrane which gives the maximum
signal-to-noise ratio. The hair bundles in this optimal condition
turn out to be in the region where amplitude death is seen, which
indicates the hair bundles are likely to exploit amplitude death for
signal transmission.

\section{ Physical model for elastically coupled hair bundles
through a massive membrane}

 We consider dynamical properties of bullfrog hair cell
bundles coupled by an overlying membrane with finite mass.
 We model the membrane by $N\times
N$ pieces of mass $m$ which are elastically coupled to each other
and also attached to hair bundles (see Fig. 1 (a)).  The equation of
motion reads
\begin{eqnarray}
\label{Seq}
 m \ddot{S_{i,j} } &=& -m\gamma
\dot{S_{i,j}}+k(S_{i+1,j}-2S_{i,j}+S_{i-1,j}) \\ \nonumber &+&
k(S_{i,j+1}-2S_{i,j}+S_{i,j-1}) + \sum_{l}f_{{\rm HB},l}
\delta_{i,I(l)}\delta_{j,J(l)}\\ \nonumber  &+&
\sum_{l}(f_{N,l}(t)+F(t)) \delta_{i,I(l)}\delta_{j,J(l)},
\\\lambda\dot{ X_{ l}}&=&-f_{{\rm HB}, l}-k_{\rm gs}({ X}_{ l}-{
X}_{{\rm a},l}-DP_{{\rm o},l}) -k_{{\rm sp},l}{ X}_{ l},
\\
\lambda_{\rm a}\dot{X_{{\rm a},l} }&=& k_{{\rm gs}}(X_{l}-X_{{\rm
a},l}-DP_{{\rm o},l})+F_{{\rm a},l}.
 \label{Xeq}
\end{eqnarray}
The parameters we used are described in Table 1.
 Here we assume  the membrane moves unidirectionally along i-axis
and $S_{ij}$ is the displacement in this direction from its
reference point.
 $X_{l}$ is the $l$-th hair bundle's deflection and it is assumed to be tied to
the mass at $(i,j)=(I(l),J(l))$\footnote{The coordinates of the
$l$-th hair cell bundles are $I(l)=(2l-1){\rm mod}N+{\rm
floor}(\frac{2l-1}{N}){\rm mod}(2), J(l)= {\rm
floor}(\frac{2l-1}{N})+1 $. },
 so that $ S_{I(l),J(l)}=X_{ l}$.  $f_{{\rm HB},l}$ is determined through Eq.(2.2) which is the force
 on the mass exerted by the $l$-th hair bundle. $f_{N,l}$ is the
 thermal noise force exerted on the $l$-th hair bundle.
$\gamma$ is the friction constant per mass of the membrane, $k$ is
the inter-bundle elastic coupling strength, and $\lambda$ is the
friction constant of a free-standing hair bundle. $\delta_{i,j}$ is
the Kronecker delta which is 1 if $i = j$, otherwise 0.
 The acoustic stimulus
 delivered to the hair bundles is $F(t)$.  The individual hair
bundles are based on the rigorous model for bullfrog sacculus hair
cells\cite{martin2000,nadrowski}. In this model, active hair-bundle
movement
 results from the ${\rm Ca}^{2+}$-dependent activity of the molecular motors, which are
 connected to transduction ion channels through gate springs.
$k_{{\rm sp},l}$ is the intrinsic stiffness of the pivot spring of
$l$-th hair bundle. While the bundle's position $X_{l}$ is set to
zero when the pivot spring force vanishes, its actual equilibrium
position can be about -70 nm due to the gating spring.
Eq.(2.3) describes the molecular motor position $X_{{\rm a},l}$ of
$l$-th hair bundle where $\lambda_{\rm a}$ is the parameter relating
a force to the velocity of the molecular motor.  $F_{{\rm
a},l}=0.14f_{{\rm max}, l} (1-0.65 P_{{\rm o},l})$ is the active
force from the molecular motor. $D$ is the gating spring elongation
and $P_{{\rm o},l}=1/(1+\exp(\frac{X_{l}-X_{{\rm a},l}+16.7{\rm
nm}}{4.53{\rm nm}}))$ is the open probability of the $l$-th ion
channel. To simulate the inhomogeneity of hair bundles, we used the
non-uniform distribution of pivot spring stiffness $k_{\rm sp}$ and
the maximal force of the motor molecules $f_{\rm max}$, where these
parameters are influenced by the minute difference in the size of
the hair bundles.
 Using Gaussian random number generators (gasdev function in {\em Numerical Recipes in
Fortran}\cite{recipes}), we generated  random
distributions of parameters around $<f_{\rm max}>=342$ pN and
$<k_{\rm sp}>=0.65$ pN/nm with a variance of about 7 pN and 0.05
pN/nm, respectively.  Fig. 1 (b) and (c) shows an example of the set
of parameters and their spatial distribution.  The experimental
error of the parameters in Martin {\it et al.} \cite{martin2000} can
be considered as the upper bound of the actual variance of the
parameters so we have chosen smaller values than the experimental
errors. This is not harmful to proving the existence of the
amplitude death as the phenomenon arises more easily for larger
variance.

For most active biological systems, accurate treatment of the
thermal noise force is important, because the coherence of
spontaneous motion is often vulnerable to thermal noise. The
autocorrelation function of the thermal noise force is usually
expressed using the white noise approximation,
$<f_{N}(t)f_{N}(t+\tau)>~ \propto ~\delta(\tau)$ ($\delta(\tau)$ is
Dirac delta function which is defined to be zero except at $\tau=0$,
and $\int d\tau \delta (\tau) = 1$). The noise force strength is
determined to satisfy the equipartition theorem, so the
proportionality constant is approximated by $2 k_{B}T$ times the
frictional constant.
 In reality, however, the correlation time is not exactly zero, so it is not possible to have the exact Dirac delta
 function.
  For the system with a
finite correlation time, we have to re-determine the strength of the
noise force; otherwise, the thermal noise does not satisfy the
equipartition theorem, $<{\rm velocity}^{2}>=k_{B}T/m$, where $T$ is
the temperature. Thus, we derive and use a relation between the
noise force strength and its correlation time satisfying the
equipartition theorem. The auto correlation function of the thermal
noise force\footnote{ We simulate the noise force by running random
variable $g_{k,l}$ in the form of $f_{N,l}(t)\propto
\sum_{k}g_{k,l}\exp(-(\frac{\sqrt{2}t}{\tau_{c}}-k-\frac{1}{2})^{2})$
where $<g_{k,l}>=0$ and $<g_{k,l}^{2}>=1$.} then reads
\begin{eqnarray}
<f_{N,l}(t)f_{N,l^{\prime}}(t+\tau)>=\frac{2k_{B}T\lambda_{\Sigma}\delta_{l,l^{\prime}}}{e^{\frac{\tau_{c}^{2}}{4}(\frac{\lambda_{\Sigma}}{m})^{2}}{\rm
erfc}(\frac{\tau_{c}}{2}(\frac{\lambda_{\Sigma}}{m}))}\frac{1}{\sqrt{\pi}\tau_{c}}e^{-(\frac{\tau}{\tau_{c}})^{2}},
\label{noisecorr}
\end{eqnarray}
where $\lambda_{\Sigma}=\lambda+m\gamma$ is the net friction
constant for individual hair bundle. One can note that the magnitude
of noise force has roughly $\sim 1/\sqrt{\tau_{c}}$ dependence on
the correlation time for $\tau=0$.
 Using
Eq.(\ref{noisecorr}), we determine the correct amplitude of the
noise force which satisfies the equipartition theorem (see
electronic supplementary material for the derivation of the
formula).

Some further remarks on our model are due. The coupling in Eq.(2.1)
describes the elastic coupling for small amplitude oscillation, and
the motion of hair bundles is assumed to be unidirectional to mimic
the fact that the bundles' deflection is mainly in the directions
towards the tallest or the shortest bundle. While we find the
amplitude death phenomenon also in a purely one-dimensional array,
we choose the particular two-dimensional array shown in Fig. 1 (a)
to describe the structure of the hair bundle array in a sacculus(see
e.g. Fig.3 of \cite{strimbu2010}). To describe the internal dynamics
of the membrane, we consider an additional mass element between two
hair bundles. So we use 100($N$=10) mass elements for the overlying
membrane while 50 hair cells are beneath the membrane. We simulate
50 hair cells for computational efficiency, even though the number
of hair cells in a frog's sacculus is about ten times
larger\cite{strimbu2010}.
 Since we do not find any significant
difference from 10 hair cells concerning the amplitude death, we do
not think the size of the array is an important factor to the
amplitude death.
 We
used a closed boundary condition as in biological systems.
 The
effective mass $m$ includes fluid compartment and otolithic mass
that are assumed to move in unison with the hair bundles.
Unfortunately, there is no experimentally known value for the
effective mass due to the difficulty of its measurement. In our
simulations, we use 2 $\mu$g for $m$ which will be shown here to be
close to an optimal value for high signal-to-noise ratio. Since a
frog sacculus is a detector for low frequency vibration and sound of
wavelengths larger than the size of the sacculus, we assume that the
acoustic stimulus is uniform across all units.

\begin{table}
\caption{List of the parameters for the simulations.  When other
values for the parameters are used, they are listed in the figures
or figure captions. }
 \centering
\begin{tabular}{cccc}
  Parameter & Definition & Value & Ref.\\ \hline
 \hline
 $m$ & the mass of one unit of crosscut membrane & 2 $\mu$g & \\
 $\tau_{c}$ & thermal noise correlation time & 1.4 ms & \\
 $k$ & inter-bundle elastic coupling stiffness & 0$\sim$ 2 pN/nm  &  \\
 $\gamma$ & friction constant per mass of the membrane & 0.5 kHz & \\
 $k_{gs}$ & gating-spring stiffness & 0.75 pN/nm  & \cite{martin2000} \\
 $D$ & gating spring elongation & 60.9 nm &\cite{martin2000}  \\
$<{\rm k}_{\rm sp}>$ & mean value of hair bundle pivot stiffness & 0.65 pN/nm & \cite{martin2000}\\
$\delta{{\rm k}_{\rm sp}}$ & variance of hair bundle pivot stiffness & 0.05pN/nm& \\
$<f_{\rm max}>$ & mean value of maximal motor force  & 342 pN  & \cite{nadrowski}\\
$\delta{f_{\rm max}}$ & variance of maximal motor force  & 7 pN & \\
  $\lambda$    &  friction of a hair bundle  & 2.8 $\mu$N s $m^{-1}$ & \cite{nadrowski} \\
  $\lambda_{a}$ & friction of adaptation motors  & 10 $\mu$N s $m^{-1}$ &\cite{hacohen}\\
  $T$  & temperature   & 300 K & \\
$P_{{\rm o},l}$  & open probability of the $l$-th ion channel  &
$1/(1+\exp(\frac{X_{l}-X_{{\rm a},l}+16.7{\rm nm}}{4.53{\rm nm}}))$  & \cite{nadrowski}\cite{jacobs}  \\
$N$ & linear size of the two-dimensional mass array & 10 & \\
 $N_{c}$ & number of hair cells & 50 & \\
  \hline
\end{tabular}
\end{table}

\section{\bf Results and Discussion}

\subsection{ Amplitude death of coupled hair bundles}

When the non-identical hair bundles are coupled elastically with
sufficient strength, we find their spontaneous oscillations become
quenched at a certain strength in the absence of thermal noise. (See
Fig. 1 (d) and 1 (e).) While the value for the coupling strength
varies depending on the distribution of the hair cell parameters, we
find the amplitude death arises around $k \sim$ 1 pN/nm.
The origin of amplitude death lies in the stabilization of the fixed
point (quiescent bundle state) as a consequence of interaction. It
can occur in coupled oscillators with parameter
mismatch\cite{aronson}, time-delayed coupling\cite{reddy}, and
nonlinear coupling\cite{prasad}. The amplitude death phenomenon of
coupled hair bundles in our simulation arises as a result of
parameter mismatches. The existence of this phenomenon is evidenced
by the recent experimental observation of the cessation of
spontaneous oscillation of hair bundles in bullfrog sacculi when
they are coupled by an overlying membrane\cite{strimbu2010}.

\begin{figure}
\centerline{\includegraphics[width=0.5\textwidth]{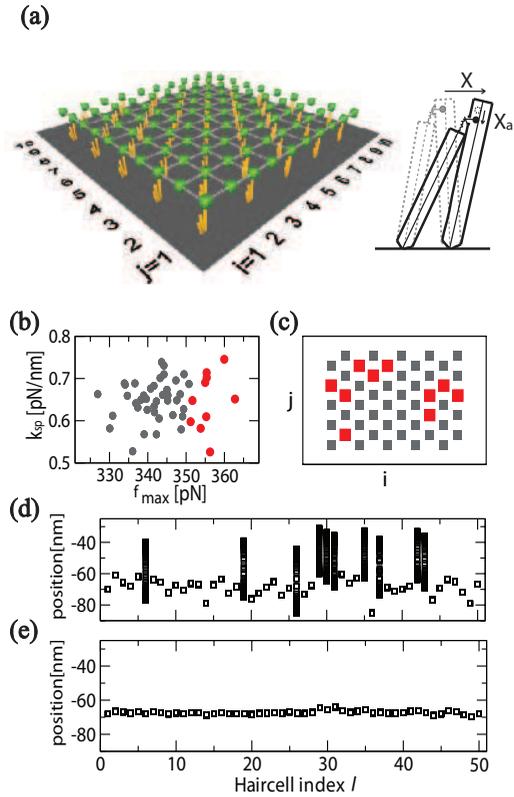}}
\caption{ (a) Schematic figure for the elastically coupled hair
bundles with mechanical loading and a schematic figure for a
free-standing hair bundle. (b) An example of the parameter
distribution. The symbols in red denote the parameters for the hair
bundles which have spontaneous oscillating motion in their
free-standing states. (c) Spatial distribution of the spontaneously
moving hair bundles (red) and quiescent bundles (grey). (d) The
stroboscopic view (snap shots at every 0.5 sec for 100 seconds) of
50 uncoupled hair bundles ($k=0$). Each hair cell is numbered using
hair cell index $l$. The position of the oscillating hair bundles
are spread out as their centre frequency of their motion is about 5
Hz. (e) The stroboscopic view of amplitude death state ($k=2$
pN/nm), which shows the cessation of the spontaneous oscillation
shown in (d). Parameters used are listed in Table 1. }
\end{figure}

The amplitude death region in the presence of thermal noise shows,
 instead of perfect quenching, suppressed fluctuations of
mechanical motions (see Fig. 2(a)). We performed numerical
simulations using 17 sets of the parameters with the same mean
values and variances. Even though we show only four of them in Fig.
2 (a) for better visibility of the figure, we found amplitude death
in all cases (see electronic supplementary material, Fig. S8).
 To examine the occurrence of the amplitude death, it proves useful
to plot the positional variance $\delta X$ instead of its Fourier
component, because the spontaneous motion is not very sinusoidal. As
inter-bundle coupling strength increases, the positional variance
$\delta X$ increases rapidly due to synchronized spontaneous
movement. It decreases later due to the amplitude death. The
amplitude death arises around k=1$pN/nm$ (See also Fig. S8). This
cross-over value for the amplitude death is not universal, but it
depends on the distribution of the parameters, the membrane mass or
the noise correlation time. We find the cross-over to the amplitude
death region arises at weaker coupling strength for shorter thermal
 correlation time $\tau_{c}$ (see Fig. S5 (a)).

 \subsection { Response of open channel probability to external
stimulus}

Let us consider the neural response of the hair bundles in the
amplitude death state.
 The influx of cations
through the transduction channels depolarizes the cell membrane
which opens voltage-gated channels at the base of the hair cell and
generates a synaptic current\cite{hudspethnature}. The information
on the auditory stimulus is passed along the auditory nerve in the
form of a spike train. Simplifying the process, we assume the neuron
spike rate is proportional to transduction current\cite{camalet}.
Then, concerning the spike rates of a bundle of the neurons, rather
than the averaged displacement of hair bundles, it is more
appropriate to consider an averaged open probability,
\begin{eqnarray}
 P^{*}_{o}=\frac{1}{N_{c}}\sum_{l=1}^{N_{c}}P_{o,l},
\end{eqnarray}
 where $N_{c}$ is the number of hair cells.

In Figs. 2 (b), (c), and (d), we plot $P^{*}_{o}(t)$ when a pure
tone stimulus at the frequency 6 Hz is applied when 40 s < time < 60
s. Compared to the case of uncoupled hair bundles (Fig. 2(b)), the
weakly coupled hair bundles (Fig. 2(c)) show a strong amplification
of the signal in the open probability $P^{*}_{o}$.  But this signal
amplification appears together with unwanted strong noise (see the
data in the absence of the signal in Fig. 2(c) when time < 40 s,
time> 60 s). The noisy fluctuation of the open channel probability
limits the threshold of the auditory signal, because the acoustic
signal drowned in the $P_{0}^{*}$ fluctuation cannot be encoded in
the neuronal signal.  The amplitude death resolves this problem. It
suppresses intrinsic noise which may exceed the input signal (see
Fig. 2(d)). It is interesting to note that the background noise of
the open probability is much weaker in the amplitude death state
(Fig. 2 (d)) than in the uncoupled system (Fig. 2 (b)).


\begin{figure}
\centerline{\includegraphics[width=0.5\textwidth]{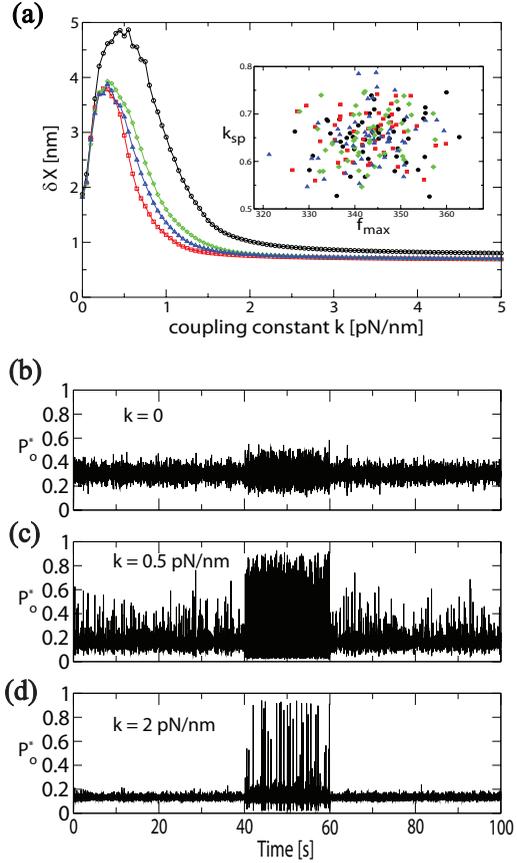}}
\caption{ (a) The positional variance $\delta X \equiv
\sqrt{\overline{X^{2}}-\overline{X}^{2}}$ of the averaged
displacement $X=\frac{1}{N_{c}}\sum_{l}X_{l}$ over hair cells where
$\overline{\cdots}$ means time-average, which shows suppression of
the mechanical fluctuation of hair bundles as the inter-bundle
coupling strength increases. The initial $\delta X$ increase is due
to the synchronization of the hair bundle movement.
 Each color denotes the
result for each set of parameters which is shown in the inset. The
response of the averaged open probability $P^{*}_{o}(t)$ to a 6 Hz
stimulus $F(t)=\sin (6\times 2\pi \frac{\rm time}{\rm sec}) $ pN for
40 s < time < 60 s (otherwise $F(t)$=0) is shown (b) when the hair
bundles are uncoupled ($k=0$), (c) when the hair bundles show a
coherent spontaneous motion ($k=0.5$ pN/nm), and
 (d) when the hair bundles are in the amplitude death region ($k=2$ pN/nm).
In the amplitude death state (d), the spontaneous fluctuation in (c)
is strongly suppressed but still the response is significantly
stronger than the uncoupled case in (b). Parameters used are listed
in Table 1.
 }
\end{figure}

\subsection{ Signal-to-noise ratio }

 The sources of noise can be divided into those
which arise from (i) fluctuation associated with an active force
generated by hair bundles, (ii) thermal fluctuation associated with
the Browninan motion of hair bundles, and (iii) fluctuation
associated with the stochastic nature of channel opening.
Source (i) appears to dominate at weak couplings where the bundles
are free-standing or showing collective spontaneous motion. Source
(ii) is the main source of noise in the amplitude death region where
the hair bundles' spontaneous movements are suppressed.

The hair bundles' active forces magnify the mechanical response of
the oscillatory stimulus, but this amplification also enhances the
background noise. To investigate the competition between the signal
amplification and the noise reduction, we calculate the power
spectra of mechanical displacement,
\begin{eqnarray}
\tilde{S}_{X}(\omega)=\frac{1}{N_{c}}\sum_{l=1}^{N_{c}}|\tilde{X}_{l}(\omega)|^{2},~~\tilde{X}_{l}(\omega)=\frac{1}{
T_{a}}\int_{0}^{ T_{a}}X_{l}(t)e^{i\omega t}dt,
\end{eqnarray}
where $T_{a}$ is the time period for the Fourier transformation and
$\omega=2\pi f$ is the angular frequency. We plot
$\tilde{S}_{X}(\omega)$ for $f=$6 Hz pure tone signal of the
amplitude 0.05 pN (Fig. 3 (a), (b), and (c)) and 0.5 pN (Fig. 3 (d),
(e), and (f)). When a weak acoustic signal is applied to uncoupled
hair bundles, the signal can be drowned in the power spectrum as
shown in Fig. 3 (a). While one can see an amplified signal by weakly
coupled hair bundles in Fig. 3 (b), the intrinsic noise is also
strong and the single tone signal at 6 Hz is vaguely seen ( see
green arrow in Fig. 3 (b)). In contrast to these cases, the input
acoustic signal can be clearly seen in the amplitude death region
(Fig. 3 (c)) where the background noise level is severely reduced by
about two orders of magnitude.

 For the signal with sufficient strengths
 (Fig. 3 (d), (e), and (f)), the amplification of the oscillatory stimulus is more prominent
 than the background noise reduction.
For the signal with the amplitude of $F= 0.5$ pN, the power spectra
of coupled hair bundles in Fig. 3(e) and (f) show the second
harmonic at $2f=$12 Hz due to the nonlinearity. This is in contrast to the case of uncoupled
hair bundles (Fig. 3 (d)) which does not show the harmonics.
When a weak signal is applied to the system in the amplitude death
region, noise reduction (rather than signal amplification) more
significantly contributes to the enhancement of the signal
transmission ( Fig. 3 (a), (b), and (c)). Provided the neuronal
threshold is low enough, the amplitude death phenomenon allows the
auditory systems to have a low threshold of acoustic stimulus.

\begin{figure}
\centerline{\includegraphics[width=0.8\textwidth]{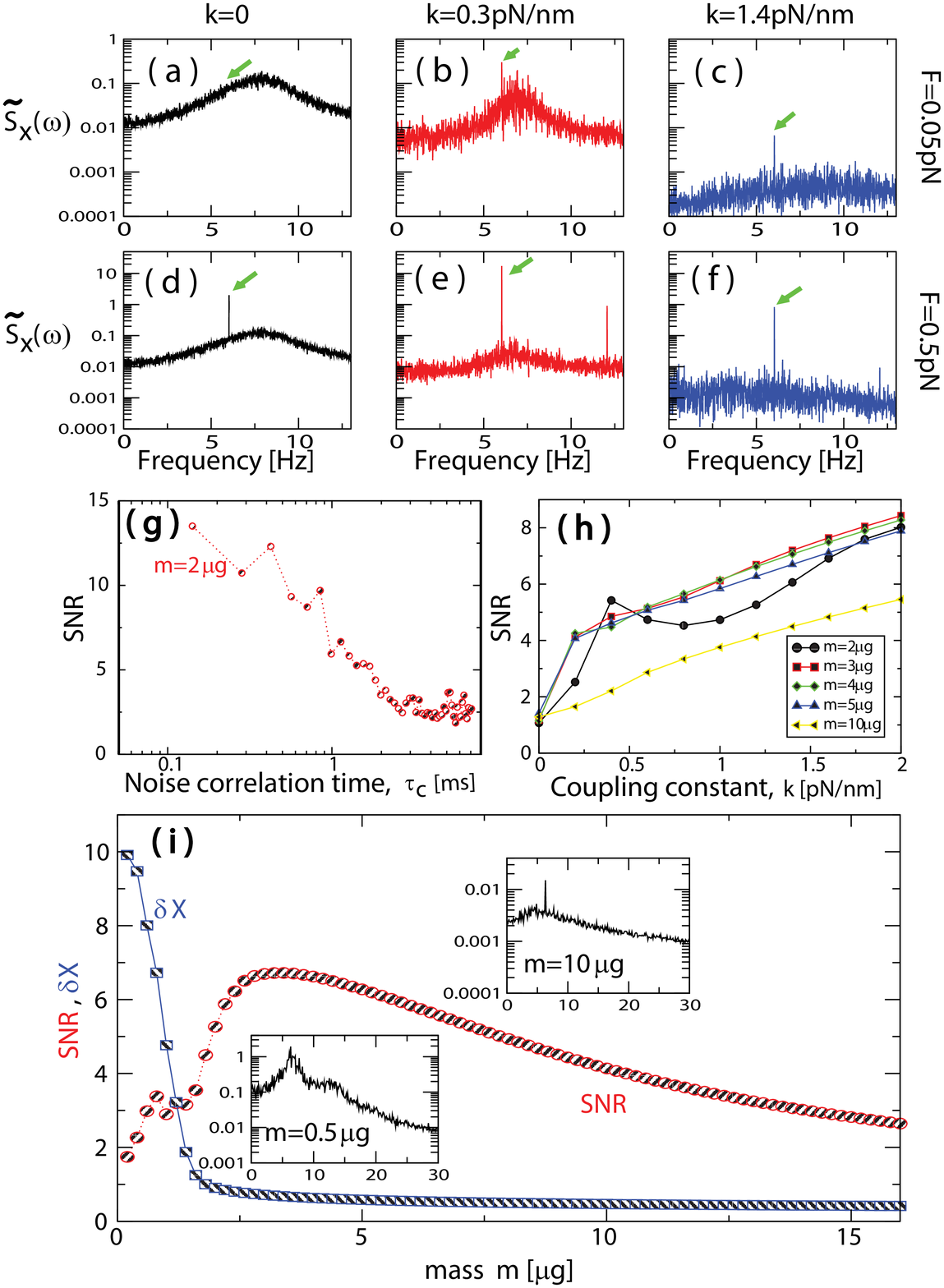}}
\caption{ The power spectra of mechanical displacement
$\tilde{S}_{X}(\omega)$ as a function of frequency $f=\omega/2\pi$
when a 6 Hz stimulus is applied with amplitude of $F$=0.05 pN
((a),(b),(c)) and $F$= 0.5 pN ((d),(e),(f)).  The coupling strengths
are ((a),(d)) $k$=0, ((b),(e)) $k$=0.3 pN/nm, ((c),(f)) $k$=1.4
pN/nm. The time period for Fourier transformation $T_{a}$ is 100s.
Note that the remarkable change in the noise floor level in the
figures from (a) to (f).
 A second harmonic at 12 Hz appears in (e) and (f). (g) SNR as a function of the
correlation time $\tau_{c}$ for $k=$1.2 pN/nm. (h) SNR as a function
of the coupling strength $k$ for the various masses and
$\tau_{c}$=1.4 ms. (i) SNR and $\delta X$ as a function of the mass
$m$ for $\tau_{c}=1.4$ ms and $k=$1.2 pN/nm, which shows that SNR
has maximum value at $m=3\mu$g in the amplitude death region. The
inset shows the power spectra for different mass $m$. For the
simulation from (g) to (i), $F=0.1$ pN.
 We used the averaged
power spectra over ten different trials of thermal noise force.
Parameters used are listed in Table 1. }
\end{figure}

Signal-to-noise ratio (SNR) is a measure that compares the level of
a desired signal to the level of background noise. It is defined by
\begin{eqnarray}
SNR=\lim_{\Delta \omega \rightarrow
0}\frac{\tilde{S}_{X}(\Omega)}{\frac{1}{\Delta
\omega}\int_{\Omega-\Delta \omega/2}^{\Omega+\Delta
\omega/2}\tilde{S}_{X}(\omega)d\omega},
\end{eqnarray}
where $\Omega$ is the angular frequency of the oscillatory stimulus;
$F(t)=F\sin\Omega t$ in Eq. (2.1). In Fig. 3 (g),(h), and (i), we
plot the SNR for various parameters. We find the SNR increases
steadily as we decrease the noise correlation time (Fig. 3 (g)).
Since the noise strength and the bandwidth of the thermal noise is
inversely proportional to the correlation time, our finding
indicates  a positive role of noise in the sense that the SNR
increases with the noise strength. However, the physical origin of
this phenomenon must be different from the orthodox theory of
stochastic resonance\cite{hanggi}, because we observed SNR is simply
decreased by a fictitious increase of the noise strength at a fixed
correlation time.


\subsection{ Optimal loading for SNR and amplitude death}

Our simulation of the system shows that the SNR has approximately
monotonic dependence
on elastic coupling strength $k$ (Fig. 3 (h)).
Meanwhile,
we find SNR has non-monotonic dependence on the mass of the
overlying membrane (Fig. 3(i)). We find that there exists an optimal
value of the mass which maximizes SNR. This is a result of the
competition between active amplification and noise reduction. The
force signal to a hair bundle is more amplified (accelerated) for
lighter masses but the hair bundle's motion also becomes more noisy.
This noisy motion is suppressed by the inter-bundle coupling and
 we find that the maximal SNR arises when the hair bundles are in the amplitude
death region. Fig. 3(i) shows when SNR reaches its peak value at
$m=3\mu$g, the positional variance $\delta X$ is already minimal.
Therefore, we conclude that the active hair bundles of a bullfrog's
sacculus may rely on the amplitude death mechanism to enhance signal
transmission.

There has been a long-standing problem on the anomalously low
threshold of acoustic stimulus in hair bundles\cite{bialek}. If the
individual hair bundle is described by a mass $m$ on a spring of
stiffness $\kappa$, its positional fluctuation in thermal
equilibrium is $\delta x =\sqrt{k_{B}T/\kappa}$ according to the
equipartition theorem. Thus, theoretical $\delta x$ is larger than 2
nm because in no inner-ear organ has the bundle stiffness been found
to exceed $\kappa \sim 1$ pN/nm\cite{bialek}. But many experimental
data suggest that the displacement of smaller than 0.1 nm is readily
detectable ( see Bialek\cite{bialek} and references therein). We
think that the amplitude death may provide a key to this problem. In
the amplitude death region, the positional fluctuations of coupled
hair bundles are strongly suppressed, so that $\delta x$ can be as
small as 0.1 nm in spite of thermal noise. In this region, the hair
bundles can detect even $F \sim 0.05$ pN where the signal is clearly
visible compared to background noise level (Fig. 3(c)). This signal
would be completely drowned out if they were uncoupled (Fig. 3(a)).
Assuming $F \sim 0.05$ pN is the minimum detectable stimulus, we can
estimate the mass per hair bundle as $m \approx 0.05 {\rm pN}/a_{\rm
th}$ where $a_{\rm th}$ is the smallest detectable peak acceleration
of vibration. Lewis {\it et al.}\cite{lewis01} reported that a
frog's sacculus can detect the acceleration down to $a_{\rm th}\sim
10^{-6}g \sim 10^{-5} {\rm m/s^{2}}$. From this value, we estimate
$m\approx 5 \mu$g which is not significantly different from the mass
causing the maximal SNR in Fig. 3(h). This estimation leads us to
reason that the anomalously low stimulus threshold $\delta x \sim$
0.1 nm might be achieved through the amplitude death mechanism,
which is difficult to be detected by an individual hair
bundle\cite{martin2000}.

\subsection{ Frequency selectivity}

\begin{figure}
\centerline{\includegraphics[width=0.8\textwidth]{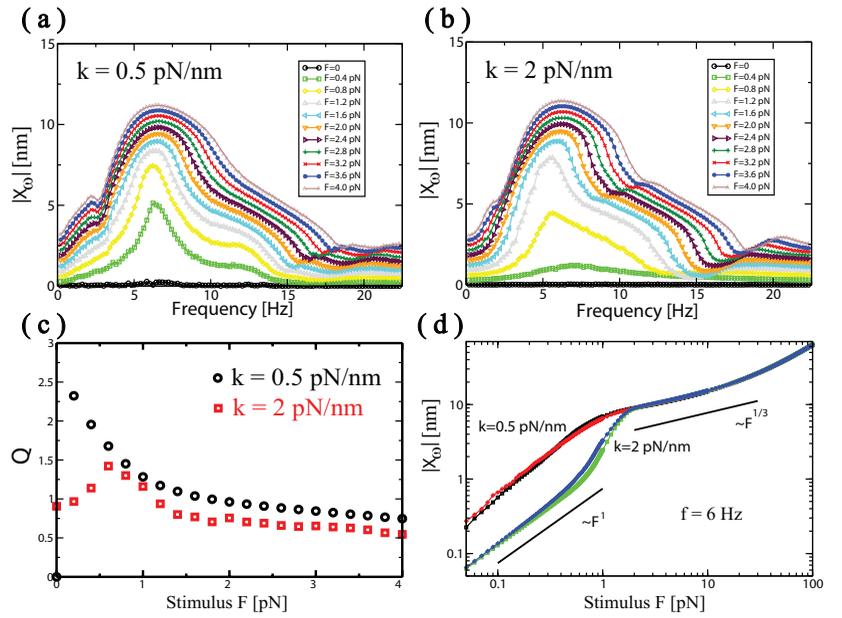}}
\caption{ (a) The response of the Fourier component of mechanical
displacement $|X_{\omega}|$=
$|\frac{1}{N_{c}}\sum_{l}\tilde{X}_{l}(\omega)|$ as a function of
the frequency $f=\omega/2\pi$ for the system in the spontaneous
oscillating region ($k=0.5$ pN/nm).  (b) The same plot for the
amplitude death region ($k= 2$ pN/nm). (c) The quality factor Q vs.
the stimulus strength F. The quality factor of the coupled bundles
is lower in the amplitude death state and weakly dependent on the
stimulus strength. (d) $|X_{\omega}|$ as a function of the stimulus
amplitude $F$ for two different coupling strengths and two different
sets of the hair-bundle-distribution parameters (denoted by
different colors). It shows the linear response for the weak
stimulus. Beyond this region, the response roughly shows a $1/3$ law
where the responses for two different coupling strengths coincide.
Parameters used are listed in Table 1. }
\end{figure}


In Fig. 4 (a) and (b), we plot the mechanical response of the hair
bundles $|X(\omega)|=|\frac{1}{N_{c}}\sum_{l}\tilde{X}_{l}(\omega)|$
for the stimulus of the frequency $f=\omega/2\pi$.
 The spontaneous oscillating region (Fig. 4(a)) shows slightly
better frequency selectivity than the amplitude death region (Fig.
4(b)) only for weak enough stimuli ($F \leq 0.8$ pN).  For the
stimuli with $F>$ 0.8 pN, the broadening of the response spectra in
both regions is similar. There is no significant difference in the
frequency selectivity between the two regions because of the
inhomogeneous distribution of hair bundles. The inhomogeneity in our
calculation was sufficient to prevent the hair bundles from locking
to a single-frequency mode. The frequency selectivity is quantified
in terms of height $|X_{\omega_{0}}|$ of the spectral peak at its
characteristic angular frequency $\omega_{0}$, and its quality
factor Q=$\omega_{0}/|\omega_{2}-\omega_{1}|$ where the halfwidth
$|\omega_{2}-\omega_{1}|$ is defined by
$|X_{\omega_{1}}|=|X_{\omega_{2}}|=|X_{\omega_{0}}|/2$. Our
numerical results for coupled hair bundles of a frog's sacculus show
Q $\sim 1$ (see Fig. 4 (c)).
 This is consistent with electrophysiological recordings from frog
auditory nerve fibers which indicate a high sensitivity to small
stimuli ($\sim $ 0.1 nm) but poor frequency selectivity (Q <
1)\cite{narins,lewis1988}. A recent experiment\cite{strimbu2012}
even reports the absence of the frequency tuning  in frog sacculi
when hair bundles are coupled to an overlying membrane. The complete
disappearance of the frequency selectivity in Strimbu {\it et
al.}\cite{strimbu2012}, however, is probably caused by the heavier
weight of the artificial membrane in the experiment. In our
calculations, as we increase the membrane mass, we observe the peak
frequency is lowered and eventually the frequency selectivity
disappears (see electronic supplementary material, Fig. S4).


\subsection{ Comparison to existing models and experiments}

Our theory as well as the lower frequency selectivity in
experiments\cite{narins,lewis1988,strimbu2012} are in contrast to
the theoretical prediction in Dierkes {\it et al.}\cite{dierkes}.
The coupled hair bundle model\cite{dierkes} shows sharp frequency
selectivity but amplitude death did not appear in the model. We
think one of the reasons for the discrepancy is that their
non-uniformity was not sufficient to cause amplitude death. It might
also be useful to note that when characteristic frequencies are
similar, coupling oscillatory and quiescent hair bundles can cause
amplitude death more efficiently.

Egu\'iluz et. al.\cite{eguiluz} and Camalet et. al.\cite{camalet}
proposed a generic model which might underlie the essential
nonlinearity\cite{rhode1971}, where hair cells are assumed to
operate at the critical point of Hopf bifurcation\cite{strogatz}.
According to the critical oscillator model\cite{eguiluz,camalet},
the amplitude response of hair bundles to an oscillatory stimulus
with strength $F$ scales as $\sim F^{1/3}$ for arbitrarily weak
signals. In our model, the hair cell bundles in the amplitude death
region show a linear response at weak stimuli. The response evolves
to the region where it shows compressive rates close to a 1/3 law
(see Fig. 4(d)). The crossover of the growth from linear to 1/3
nonlinear rates in Fig. 4(d) is consistent with the analytic formula
derived from the generic mathematical model of amplitude death (see
electronic supplementary material). The hair-bundle nonlinearity
shown here is similar to what has been consistently observed in
experiments on mammalian cochlea after Rhode\cite{rhode1971}, where
the transition between linear and compressive nonlinear growth has
been reported\cite{ruggero97,ruggero00,nuttal1996}. Our theory is
not incompatible with the theory\cite{camalet,nadrowski} of
individual hair bundles. As in Nadrowski {\it et
al.}\cite{nadrowski}, our individual hair bundles are assumed to be
near the Hopf bifurcation critical point. They were randomly
distributed near the critical point with a small width of the
distribution.

 Stochastic resonance is a phenomenon where SNR is maximized
in the presence of a specific non-zero level noise. This phenomenon
widely occurs in  threshold systems, such as a two-state system
separated by an energy barrier. It has been speculated that in the
case of free-standing hair bundles (such as inner hair cell
bundles), the noise plays a role in acoustic signal processing, by
enhancing the sensitivity of the system through stochastic
resonance\cite{jaramillo}. While it has been reported that applying
noise to sacculi may enhance the SNR\cite{indresano}, it is
difficult to judge whether the sensation is indeed improved by
stochastic resonance.
 One of the reasons is that the generic stochastic resonance model is not applicable to
a coupled system with active dynamics. Instead we ask whether the
noise plays any positive role in signal transmission. We find out
that the direct increase of the noise strength (e.g. by increasing
temperature) at a fixed $\tau_{c}$ simply decreases the SNR. In this
sense, the orthodox stochastic resonance phenomenon does not occur
in the system. Meanwhile, we may say that the noise can have a
positive role in signal transmission, in the sense that the thermal
noise enhances the SNR by suppressing the spontaneous activity of
hair bundles. To explain the process, we need to mention that there
are two different types of apparently similar hair bundles moving in
the presence of noise. The first type spontaneously oscillates
regardless of noise. The second type is quiescent in the absence of
the noise, but they are in motion in the presence of noise. As the
thermal correlation time becomes shorter, the motion of the second
type of hair bundles vanishes, thereby enhancing the SNR (see
electronic supplementary material, Fig. S3 (a)).

Barral {\it et al.}\cite{barral} performed experiments on a hair
bundle of the bullfrog sacculus which is coupled to simulated
'cyber' bundles. They argue that the coupling increases the phase
coherence between hair bundles, which endows the hair bundles with
better characteristics. This is in contrast to our suggestion that
the cessation of the oscillation has useful properties. We think the
discrepancy between the two is a question of the strength of the
coupling and the inhomogeneity of the bundles. Barral {\it et
al}\cite{barral} went up to around 0.5 pN/nm stiffness, where we
expect they could also observe amplitude death if they would
increase the coupling stiffness. Regarding the coupling strength,
how strong is strong enough for the amplitude death phenomenon
depends on the inhomogeneity of hair bundles (see electronic
supplementary material). We believe, if the two bundles are
sufficiently different (for instance one is quiescent and the other
is oscillating) then the experiments of the type in Barral {\it et
al}\cite{barral} can show amplitude death even for the coupling of
0.5 pN/nm stiffness.

\section{Conclusion}

Using numerical simulations, we have demonstrated that a signal
amplification and noise reduction in coupled hair bundles can appear
through amplitude death. The signal-to-noise ratio can be maximized
when the coupled hair bundles are in the amplitude death state.
While individual hair bundles in a frog's sacculus show spontaneous
oscillation, we think this oscillation in vivo is suppressed to
obtain high signal-to-noise ratio. Our finding is consistent with
the recent experimental observation\cite{strimbu2010} showing the
absence of the spontaneous oscillations when they are coupled with
an overlying membrane.

When individual systems are coupled, the net system is often
described as coupled individual systems or sometimes is considered
as one new system. In which class, do our hair bundles in the
amplitude death region belong to?
 The minimum inter-bundle coupling strengths  for the
amplitude death in most of our calculations were comparable to the
intrinsic stiffness of hair bundles ($k_{\rm sp}=$0.65 pN/nm). Since
the amplitude death phenomenon can arise even when the inter-bundle
coupling is weaker than the intrinsic stiffness, we think that the
hair bundles in amplitude death region can be considered as weakly
coupled independent oscillators although more quantitative analysis
is necessary to clarify this issue.


 The transduction mechanism presented in this work can be
used to fabricate low-noise amplification in acoustic sensors. A
cantilever-type mechanical sensor can act similar to hair cell
bundle when it has biomimetic feedback action\cite{kim,song}. The
amplitude death phenomenon in a coupled array of these biomimetic
sensors will lead to a low-noise amplification with an enhanced
signal-to-noise ratio.

\ack{
 This work was supported by Basic
Science Research Program through the National Research Foundation of
Korea (NRF) funded by the MEST (2011-0009557). The author thanks
Marie Curie Fellowship
 (2010-2011) of EU  through University of Bath in UK.
}

\newpage

\begin{center}{\bf Figure Captions}
\end{center}

\vspace{0.5cm}
 {\bf Fig. 1}
(a) Schematic figure for the elastically coupled hair bundles with
mechanical loading and a schematic figure for a free-standing hair
bundle. (b) An example of the parameter distribution. The symbols in
red denote the parameters for the hair bundles which have
spontaneous oscillating motion in their free-standing states. (c)
Spatial distribution of the spontaneously moving hair bundles (red)
and quiescent bundles (grey). (d) The stroboscopic view (snap shots
at every 0.5 sec for 100 seconds) of 50 uncoupled hair bundles
($k=0$). Each hair cell is numbered using hair cell index $l$. The
position of the oscillating hair bundles are spread out as their
centre frequency of their motion is about 5 Hz. (e) The stroboscopic
view of amplitude death state ($k=2$ pN/nm), which shows the
cessation of the spontaneous oscillation shown in (d). Parameters
used are listed in Table 1.

\vspace{0.5cm}
 {\bf Fig. 2}
(a) The positional variance $\delta X \equiv
\sqrt{\overline{X^{2}}-\overline{X}^{2}}$ of the averaged
displacement $X=\frac{1}{N_{c}}\sum_{l}X_{l}$ over hair cells where
$\overline{\cdots}$ means time-average, which shows suppression of
the mechanical fluctuation of hair bundles as the inter-bundle
coupling strength increases. The initial $\delta X$ increase is due
to the synchronization of the hair bundle movement.
 Each color denotes the
result for each set of parameters which is shown in the inset. The
response of the averaged open probability $P^{*}_{o}(t)$ to a 6 Hz
stimulus $F(t)=\sin (6\times 2\pi \frac{\rm time}{\rm sec}) $ pN for
40 s < time < 60 s (otherwise $F(t)$=0) is shown (b) when the hair
bundles are uncoupled ($k=0$), (c) when the hair bundles show a
coherent spontaneous motion ($k=0.5$ pN/nm), and
 (d) when the hair bundles are in the amplitude death region ($k=2$ pN/nm).
In the amplitude death state (d), the spontaneous fluctuation in (c)
is strongly suppressed but still the response is significantly
stronger than the uncoupled case in (b). Parameters used are listed
in Table 1.

\vspace{0.5cm}
 {\bf Fig. 3}
The power spectra of mechanical displacement $\tilde{S}_{X}(\omega)$
as a function of frequency $f=\omega/2\pi$ when a 6 Hz stimulus is
applied with amplitude of $F$=0.05 pN ((a),(b),(c)) and $F$= 0.5 pN
((d),(e),(f)).  The coupling strengths are ((a),(d)) $k$=0,
((b),(e)) $k$=0.3 pN/nm, ((c),(f)) $k$=1.4 pN/nm. The time period
for Fourier transformation $T_{a}$ is 100s. Note that the remarkable
change in the noise floor level in the figures from (a) to (f).
 A second harmonic at 12 Hz appears in (e) and (f). (g) SNR as a function of the
correlation time $\tau_{c}$ for $k=$1.2 pN/nm. (h) SNR as a function
of the coupling strength $k$ for the various masses and
$\tau_{c}$=1.4 ms. (i) SNR and $\delta X$ as a function of the mass
$m$ for $\tau_{c}=1.4$ ms and $k=$1.2 pN/nm, which shows that SNR
has maximum value at $m=3\mu$g in the amplitude death region. The
inset shows the power spectra for different mass $m$. For the
simulation from (g) to (i), $F=0.1$ pN.
 We used the averaged
power spectra over ten different trials of thermal noise force.
Parameters used are listed in Table 1.

 \vspace{0.5cm}
 {\bf Fig. 4}
(a) The response of the Fourier component of mechanical displacement
$|X_{\omega}|$= $|\frac{1}{N_{c}}\sum_{l}\tilde{X}_{l}(\omega)|$ as
a function of the frequency $f=\omega/2\pi$ for the system in the
spontaneous oscillating region ($k=0.5$ pN/nm).  (b) The same plot
for the amplitude death region ($k= 2$ pN/nm). (c) The quality
factor Q vs. the stimulus strength F. The quality factor of the
coupled bundles is lower in the amplitude death state and weakly
dependent on the stimulus strength. (d) $|X_{\omega}|$ as a function
of the stimulus amplitude $F$ for two different coupling strengths
and two different sets of the hair-bundle-distribution parameters
(denoted by different colors). It shows the linear response for the
weak stimulus. Beyond this region, the response roughly shows a
$1/3$ law where the responses for two different coupling strengths
coincide. Parameters used are listed in Table 1.

\end{document}